\documentstyle[12pt]{article}

\newlength{\extraspace}
\newlength{\extraspaces}
\def\bsklength{.8mm} %{2mm} % for more than double spacing
\newcommand{\beq}{\begin{equation}}
\newcommand{\eeq}{\end{equation}}

\newcommand{\beqa}{\begin{eqnarray}}
\newcommand{\eeqa}{\end{eqnarray}}

\def\half{{\textstyle{1\over 2}}}
\def\pa{\partial}

\def\CL{{\cal L}}

\def\CO{{\cal O}}
\def\CS{{\cal S}}

\renewcommand{\hat}{\widehat}
\renewcommand{\tilde}{\widetilde}

\newcommand{\e}{{\rm e}}

\newfont{\cmss}{cmss10 scaled\magstep1} 
\newfont{\cmsss}{cmss10 }% scaled 833}%at 7pt
\def\IZ{\relax\ifmmode\mathchoice
{\hbox{\cmss Z\kern-.4em Z}}{\hbox{\cmss Z\kern-.4em Z}}
{\lower.9pt\hbox{\cmsss Z\kern-.4em Z}}
{\lower1.2pt\hbox{\cmsss Z\kern-.4em Z}}\else{\cmss Z\kern-.4em Z}\fi}
\def\IR{\relax\ifmmode\mathchoice
{\hbox{\cmss I\kern-.5em I}}{\hbox{\cmss R\kern-.5em R}}
{\lower.9pt\hbox{\cmsss I\kern-.5em I}}
{\lower1.2pt\hbox{\cmsss R\kern-.5em R}}\else{\cmss R\kern-.5em R}\fi}

\def\del{\nabla}
\def\d{{\rm d}}
\newcommand{\Mp}{{m_{\rm pl}}}

\begin{document}
\setcounter{page}{0}
\addtolength{\baselineskip}{\bsklength}
\thispagestyle{empty}
\renewcommand{\thefootnote}{\fnsymbol{footnote}}	%for symbols

\begin{flushright}
{\sc MIT-CTP-2540}\\
hep-th/9606060\\
June 1996
\end{flushright}
\vspace{.4cm}

\begin{center}
{\Large
{\bf{Effective Dilaton Potential in Linearized Gravity}}}\\[1.2cm] %title
{\rm HoSeong La}%						%author
\footnote{e-mail address: hsla@mitlns.mit.edu\\			%email
This work is supported in part by funds provided by the U.S. Department of
Energy (D.O.E.) under cooperative research agreement \#DF-FC02-94ER40818.
 }\\[3mm]							
{\it Center for Theoretical Physics\\[1mm]			%address
Laboratory for Nuclear Science\\[1mm]
Massachusetts Institute of Technology\\[1mm]
Cambridge, MA 02139-4307, USA} \\[1.5cm]

%{\sc Abstract}\\[1cm]
{\parbox{14cm}{
\addtolength{\baselineskip}{\bsklength}
\noindent
Considering the linearized gravity with matter fields, the effective potential
of the ``conformal dilaton'' in the string frame is generated semiclassically
by one-loop contribution of heavy matter fields. This in turn generates a
nontrivial potential for the physical dilaton in the Einstein frame with the
trace of the graviton in the Einstein frame gauged away. The remaining manifest
local spacetime symmetry is only the volume preserving diffeomorphism 
symmetry. The consistency of this procedure is examined and the possibility of
spontaneous diffeomorphism symmetry breaking is suggested.
}
}\\[1.5cm]

{Submitted to {\it Physics Letters B}} 

\end{center}
\noindent
\vfill

%%%%%%%%%%%%%%%%%%%%%%%%%%%%%%%%%%%%%%%%

\setcounter{section}{0}
\setcounter{equation}{0}
\setcounter{footnote}{0}
\renewcommand{\thefootnote}{\arabic{footnote}}	%for numbers
\newcounter{xxx}
\setlength{\parskip}{2mm}
\addtolength{\baselineskip}{\bsklength}
\newpage

%\newsection{Introduction}

%\noindent
One of the most elusive parts of particle physics is the mystery of  scalar
particles. Introduction of spontaneous symmetry breaking induced by scalar
particles in the SU(2)$\times$U(1) gauge theory has led us to the triumphant
electroweak theory, but we are still bothered by the missing existence of such
a scalar particle. In much higher energy  scale, we have encountered another
elusive scalar particle called the dilaton, which often appears in the context
of string motivated supergravity models. The actual role of the dilaton is yet
to be fully understood, but we normally anticipate that it would determine
coupling constants near the Planck scale\cite{wittdil} and clarify the
relation between the supersymmetric structure of string theory and that of 4-d
supergravity below the compactification scale, and perhaps play a role in 
4-d supersymmetry breaking itself\cite{ds}\cite{kaplo}.
%\footnote{It is expected that the
%massive dilaton vacuum would break the spacetime  supersymmetry\cite{ds}, but
%this does not necessarily mean the 4-d spacetime supersymmetry is broken in a
%massive dilaton vacuum, although the 10-d local supersymmetry must be broken in
%such a vacuum.}.

In fact, the appearance of the dilaton is quite generic in any theory inherited 
from a scale-invariant or conformally invariant gravitational theory. We
however do not know what actually controls the dynamics of the  dilaton because
its potential is unknown. As alluded in \cite{sdiffgr}, we suspect the
difficulty might partly lie on the existence of conformal  diffeomorphisms.
From this point of view, here we shall investigate the dilaton in the context
of the linearized gravity with heavy matter fields. 

The linearized theory we deal with is not renormalizable as soon as higher
order terms are included. Thus we should accommodate the theory more or less in 
spirit of effective field 
theory\cite{wilson}\cite{weinberg}\cite{donoghue}\cite{effrev}.  This approach
is fairly reasonable as long as we remain in the scale where all the higher
order contributions are sufficiently suppressed. We also assume there is a
well-defined quantum gravity at the Planck scale, e.g. superstring theory,
whose effective  linearized gravity limit looks like the one we consider here. 
This allows us to avoid any anomalous situation that may arise. In principle,
linearization of gravity must exist for any quantum gravity if we wish to
regard the graviton as a particle because particles only make sense in a local
Lorentz frame. Under this circumstance we can also treat the local spacetime
symmetry  in an equal footing as other local internal symmetries. Then we claim
that it is possible to show that the {\it conformal dilaton}  and the graviton
behave differently\footnote{To serve our purpose the best, we shall, from here
on, call the trace of the graviton the conformal dilaton and the rest of the 
graviton just the graviton.}. In a naive sense, this amounts either an anomaly
or spontaneous diffeomorphism symmetry breaking. But the effective potential
itself is Diff (i.e. diffeomorphism) invariant, hence we could interpret it as
spontaneous Diff symmetry breaking and the conformal dilaton gets a vacuum
expectation value. Furthermore, if we accept such a symmetry breaking, we can
argue that the physical dilaton incorporating the string dilaton becomes
massive, while the graviton remains to be massless. The remaining manifest
symmetry is the SDiff (i.e. volume-preserving diffeomorphism) symmetry. Note
that this is different from the approach in \cite{veltman}, where the mass of
the graviton is allowed in equal footing. 

Before we integrate out all the matter degrees of freedom in the linearized 
gravity, the renormalizable part of the Lagrangian is only approximately
Diff-invariant, that is, up to higher orders of $\kappa$, where $1/\kappa
\equiv\Mp = 1/\sqrt{8\pi G} \sim 10^{18}$GeV. The matter part of the Lagrangian
however is exactly SDiff-invariant. Usually, there is a difficulty of
computing the semi-classical effective potential involving external gauge
fields in a gauge invariant way. However, in our case we shall only consider up
to the tadpole contribution, which turns out to be Diff-invariant.  The higher
order terms are only SDiff-invariant. Nonetheless, it is sufficient for our
purpose. 

%\newsection{Linearized Lagrangian}

%\noindent
In general, any matter-gravity couplings break the diffeomorphism symmetry
$h_{\mu\nu} \to h_{\mu\nu} +\pa_\mu \xi_\nu + \pa_\nu \xi_\mu$ in the
linearized gravity, hence we tend to think they are not allowed.
Note that this linearized diffeomorphism transformation is reduced from
the metric transformation $g_{\mu\nu} \to g_{\mu\nu} + \kappa(\del_\mu \xi_\nu
+\del_\nu \xi_\mu)$ such that
\beq
\label{eqtf1}
\delta h_{\mu\nu} = \pa_\mu \xi_\nu + \pa_\nu \xi_\mu 
+\kappa\left(\xi^\alpha\pa_\alpha h_{\mu\nu}
+h_{\mu\alpha}\pa_\nu\xi^\alpha + h_{\alpha\nu}\pa_\mu\xi^\alpha\right) \ ,
\eeq
where $g_{\mu\nu} = \eta_{\mu\nu} +\kappa h_{\mu\nu}$ and
\beq
g^{\mu\nu} =\eta^{\mu\nu} -\kappa h^{\mu\nu} + 
\kappa^2 h^{\mu\alpha}h^{\nu}_{\ \alpha} +\cdots 
\eeq
Then the above linear diffeomorphism transformation is obtained by taking
$\kappa\to 0$ limit. Similarly, a scalar $S$ transforms 
\beq
\label{eqtf2}
\delta S=\kappa\xi^\alpha\pa_\alpha S
\eeq 
and  $\delta S = 0$ is used in the linearized gravity. Thus one can easily see
that any matter-gravity couplings are not allowed. This seems to be inevitable 
if we want the linearized theory to be related to the Einstein gravity.
Nevertheless, such a missing matter-gravity coupling never conflicts with the
physics in the Newtonian limit because no individual particle can have a
significant matter-gravity coupling in this limit.  This is why, in the
Newtonian limit, matters are usually dealt in a bulk. The missing
matter-gravity coupling however becomes an issue if we need to deal with a
heavy particle. 

This difficulty however can be overcome if we require the Diff
symmetry in a more realistic way.  We are interested in the system in which
there is another mass parameter not extremely small compared to $\Mp$, yet
small enough to treat the gravity classically. So we could  require the theory
approximately Diff-invariant,  i.e. invariant up to higher orders of $\kappa$,
under  eqs.(\ref{eqtf1})(\ref{eqtf2}). Then $S + \kappa hS$ type of
matter-gravity  coupling can be introduced. 

This in turn creates another problem.  Naive truncation in the $1/\Mp$ 
expansion leaves a nonrenormalizable derivative-coupling matter-gravity 
interaction. In our case however more careful truncation drops such a term 
because the $\kappa hS$ coupling can be relatively enhanced by another big 
parameter as follows.

The system we investigate contains two massive fields: a scalar field $\varphi$
and a fermion $\psi$. If the mass $\mu$ of $\varphi$ and the fermion mass $m$  
are heavy enough, then $\mu/\Mp$ and $m/\Mp$ are not negligible compared to the
contribution of the linear graviton. It turns out that $\mu$ and $m$ need to be
of similar order to be reasonable. In this case the renormalizable scalar field
couplings include only conformal dilaton couplings. The conformal dilaton also
couples to the fermion with a Yukawa coupling constant $\tilde{\lambda}$
induced from the fermion mass as $\tilde{\lambda} \equiv m/2\Mp$. The rest of
matter-gravity couplings are all suppressed by inverse powers of the 
modified Planck mass $\Mp$. 

Thus we write the action:
\beq
\label{eq1}
\CS_{{\rm l.g.}} = \int \d^4x \CL_{{\rm l.g.}} \equiv \int d^4x 
\left(\CL_c + \CL_0 + \CL_h + \CL_\varphi + \CL_\psi  +\cdots\right),
\eeq
where
\beqa
\label{eqcos}
\CL_c &=& \Lambda_0 \left(1 + \half{\kappa h} +{\textstyle{1\over 8}}
{\kappa^2 h^2} -{\textstyle{1\over 4}}
{\kappa^2 h^{\alpha\beta}h_{\alpha\beta} } 
+ \CO({\kappa^3}) \right),
\\[1mm]
\label{eq0}
\CL_0 +\CL_h &=& {\textstyle {1\over 8}}\pa^\mu h^{\alpha\beta} \pa_\mu
h_{\alpha\beta} -{\textstyle{1\over 4}} \pa_\mu h^{\mu\alpha}
\pa^\nu h_{\nu\alpha}
 -{\textstyle{1\over 8}}\pa^\mu h \pa_\mu h
+{\textstyle{1\over 4}}\pa^\mu h \pa^\nu h_{\mu\nu}\ ,\\[1mm]
\label{eqscal}
\CL_\varphi &=& -\half\pa^\mu\varphi\pa_\mu\varphi +\half\mu^2\varphi^2
+\half\lambda_1 h\varphi^2 +\half\lambda_2 h^2\varphi^2 
-\half\lambda_\gamma h^{\alpha\beta}
h_{\alpha\beta}\varphi^2 -{\lambda\over 4}\varphi^4\  ,\\[1mm] 
&&\quad\quad\quad 
\lambda_1\equiv {\textstyle{1\over 2}{\mu^2\over\Mp}},\
\lambda_2\equiv {\textstyle{1\over 8}{\mu^2\over\Mp^2}},\ 
\lambda_\gamma\equiv {\textstyle{1\over 4}{\mu^2\over \Mp^2}}\ ,
\nonumber\\[1mm]
\label{eqfer}
\CL_\psi &=& \overline{\psi} (i\gamma^\mu\pa_\mu -m)\psi -\tilde{\lambda}
h\overline{\psi}\psi,
\quad\quad \tilde{\lambda}\equiv {\textstyle{m\over 2\Mp}}\ ,
\eeqa
where $h \equiv h^\alpha_{\ \alpha}$ and the indices are raised and lowered by
the Minkowski metric $\eta_{\mu\nu}$. The ellipsis contains nonrenormalizable
terms suppressed by the order $\CO({1\over \Mp})$. $\CL_c$ is included to
provide the necessary counter terms. This linearized action has the approximate
local gauge symmetry under eqs.(\ref{eqtf1})(\ref{eqtf2}) as well as the local
Lorentz symmetry. In particular, the action is not invariant term by term. 
%In fact, the matter kinetic energies require nonrenormalizable
%dimension-five terms to be invariant. But these terms are relatively
%suppressed by the heavy particle mass compared to the renormalizable terms

Once we decide to count the order $\kappa$ terms in 
eqs.(\ref{eqtf1})(\ref{eqtf2}), the separation of $h$ and $\gamma_{\mu\nu}
\equiv h_{\mu\nu} - {1\over 4}\eta_{\mu\nu} h$ is no longer practical because
$\delta h = 0$ is not the complete volume-preserving condition.
So we introduce a new parametrization for the conformal dilaton $\hat{h}$ and 
the graviton $\hat{\gamma}_{\mu\nu}$, incorporating order $\kappa$ terms,
to serve our purpose the best:
\beqa
\label{enewf}
\hat{h} &\equiv& h - \half\kappa h^{\alpha\beta}h_{\alpha\beta},\\
\hat{\gamma}_{\mu\nu} &\equiv& h_{\mu\nu} -{\textstyle{1\over 4}}\eta_{\mu\nu}
{h} +\kappa\left({\textstyle{1\over 32}}\eta_{\mu\nu} h^2
+{\textstyle{1\over 8}}\eta_{\mu\nu}\eta_{\mu\nu}h^{\alpha\beta}
h_{\alpha\beta} -{\textstyle{1\over 4}}h_{\mu\nu} h\right).
\eeqa
The fields now transform under Diff as\footnote{From here on, the equality 
always stands for up to the leading order of $\kappa$.}
\beqa
\label{eq2}
\delta\hat{\gamma}_{\mu\nu} &=& \pa_\mu\xi_\nu + \pa_\nu\xi_\mu
-\half\eta_{\mu\nu}\pa_\alpha\xi^\alpha 
+\kappa\left(\xi^\alpha\pa_\alpha\hat{\gamma}_{\mu\nu}
+\hat{\gamma}_{\alpha\nu}\pa_\mu\xi^\alpha
+\hat{\gamma}_{\mu\alpha}\pa_\nu\xi^\alpha 
-\half\hat{\gamma}_{\mu\nu}\pa_\alpha\xi^\alpha
\right) ,\\
\delta \hat{h} &=& 2\pa_\alpha\xi^\alpha +\kappa\xi^\alpha\pa_\alpha\hat{h} 
\eeqa
such that $(\eta^{\mu\nu} -\kappa h^{\mu\nu})\delta\hat{\gamma}_{\mu\nu} = 0$.
This gives the linearized version of the metric h-decomposition used in
\cite{sdiffgr}.
Under SDiff,
\beq
\label{esvol}
\pa_\alpha\xi^\alpha +\half\kappa\xi^\alpha\pa_\alpha\hat{h} = 0
\eeq 
so that
\beqa
\label{esdiff}
\delta\hat{\gamma}_{\mu\nu} &=& \pa_\mu\xi_\nu + \pa_\nu\xi_\mu
+\kappa\left(
\hat{\gamma}_{\alpha\nu}\pa_\mu\xi^\alpha
+\hat{\gamma}_{\mu\alpha}\pa_\nu\xi^\alpha 
+\xi^\alpha\pa_\alpha\hat{\gamma}_{\mu\nu} 
+{\textstyle{1\over 4}}\eta_{\mu\nu}\xi^\alpha\pa_\alpha \hat{h}\right) ,\\
\delta \hat{h} &=& 0.
\eeqa
Under Weyl transformations, simply
\beq
\label{eweyl}
\delta\hat{\gamma}_{\mu\nu} = 0,\quad
\delta \hat{h} = \rho.
\eeq
In this decomposition the conformal dilaton and the graviton never mix
under SDiff or Weyl.

In terms of $\hat{h}$ and $\hat{\gamma}_{\mu\nu}$
each term in the Lagrangian now reads
\beqa
\label{eqcosx}
\CL_c &=& \Lambda_0 \left(1 + \half\kappa\hat{h} +{\textstyle{1\over 8}}
\kappa^2\hat{h}^2 + \CO(\kappa^3) \right),
\\[1mm]
\label{eq0x}
\CL_0 &=& {\textstyle {1\over 8}}\pa^\mu \hat{\gamma}^{\alpha\beta} \pa_\mu
\hat{\gamma}_{\alpha\beta} 
-{\textstyle{1\over 4}} \pa_\mu\hat{\gamma}^{\mu\alpha}
\pa^\nu\hat{\gamma}_{\nu\alpha}\ ,\\[1mm]
\label{eqhx}
\CL_{\hat h} &=& -{\textstyle{3\over 64}}\pa^\mu \hat{h} \pa_\mu \hat{h}
+{\textstyle{1\over 8}}\pa^\mu \hat{h} \pa^\nu \hat{\gamma}_{\mu\nu}\ ,\\[1mm]
\label{eqscalx}
\CL_\varphi &=& -\half\pa^\mu\varphi\pa_\mu\varphi +\half\mu^2\varphi^2
+\half\lambda_1 \hat{h}\varphi^2 +\half\lambda_2 \hat{h}^2\varphi^2 
-{\lambda\over 4}\varphi^4\  ,\\[1mm] 
\label{eqferx}
\CL_\psi &=& \overline{\psi} (i\gamma^\mu\pa_\mu -m)\psi -\tilde{\lambda}
\hat{h}\overline{\psi}\psi .
\eeqa
Note that there is no renormalizable matter-graviton couplings in this
parametrization. The graviton appears only in the kinetic energy term. 

Now we would like to call the reader's attention to the fact that each term in
the matter part of the action is in fact exactly SDiff-invariant. This could
tempt us to start with the SDiff-invariance as a fundamental symmetry rather
than Diff-invariance.  Since the difference between Diff and SDiff are
conformal diffeomorphisms\cite{sdiffgr}, if the theory has conformal fixed
points where the beta function of the Newton's constant vanishes, the SDiff
symmetry will be enhanced to Diff symmetry at these points. And we can
interpret the points away from the fixed points are those with broken Diff
symmetry by a condensate or a nonperturbative effect. But this requires to
include the quantum effects of the gravity, which is beyond the scope of this
paper. 

%\newsection{Semiclassical Effective Potential}

%\noindent
We have chosen the fermion coupled to the conformal dilaton heavy enough to be
nonnegligible. Yet we want it to be still light enough compared to the Planck
scale so that we could still treat the gravity classically. So we assume there
is such an intermediate mass scale fermion in nature. Perhaps we could use any
grand-unification scale fermion. In this sense it is quite natural to include
the scalar-gravity coupling because scalar fields are to be present for the
gauge symmetry breaking. We however would not  bother any gauge field
contributions partly because their gravitational couplings are not
renormalizable and suppressed. Now the matter fields are heavy enough, thus
the conformal dilaton-matter coupling contribution is no longer negligible. We
are interested in the effect of such heavy fields in this linearized gravity.

In this intermediate regime we can safely use the semiclassical method to
integrate out the scalar and fermionic contribution to obtain the effective 
potential $V_{\rm eff} (\hat{h})$, incorporating all the necessary tree level 
contributions. This effective potential only makes sense if $h/\Mp \ll 1$ so
that only a few leading terms are meaningful. To compute the leading terms,
let us set $\varphi_c^2 = \mu^2/\lambda$ which can be taken from the gauge
symmetry breaking parameter. Then we obtain
\beqa
\label{e7}
V_{\rm eff}(\hat{h}) &=& \Lambda + a_1 \kappa\hat{h} 
+ a_2 \kappa^2\hat{h}^2 +\cdots\ ,\\
\Lambda &=& -\Lambda_0 -{\mu^4\over 4\lambda}
+{\mu^4\over 16\pi^2}\left({\rm log}{2\mu^2\over M^2} -{{3\over 2}}
 \right) -{m^4\over 16\pi^2}\left({\rm log}{m^2\over M^2} -{3\over 2}\right), 
\nonumber\\
a_1 &=&	-\half\Lambda_0 -{\lambda_1\mu^2\over 2\kappa\lambda} 
-{\mu^4\over 32\pi^2}\left({\rm log}{2\mu^2\over M^2}-1\right) 
-{m^4\over 8\pi^2}\left({\rm log}{m^2\over M^2}-1\right) ,\nonumber \\
a_2 &=& -{\textstyle{1\over 8}}\Lambda_0 
-{\lambda_2\mu^2\over 2\kappa^2\lambda} 
-{\mu^4\over 128\pi^2}\left({1\over 2}{\rm log}{2\mu^2\over M^2}-1\right) 
-{m^4\over 32\pi^2}\left(3{\rm log}{m^2\over M^2} -1\right) ,\nonumber
\eeqa
where $M$ is the renormalization scale. If we truncate at the first order term,
i.e. counting only the  tadpole contribution, $V_{\rm eff}$ is Diff-invariant.
However, if we include the quadratic term, it is no longer Diff-invariant
because $a_2\neq a_1/4$. As we pointed out before, this could be due to our
inability to compute the effective action in a Diff-invariant way.  Perhaps,
the conformal dilaton should not be really treated as a slowly varying external
field in certain energy region above the mass scale of the matter fields.
Therefore, we should really consider only the first two Diff-invariant terms.

One may wonder if there is any missing contribution from nonrenormalizable 
terms. Indeed there is and it changes numerically, but does not modify our
conclusion. In fact, the dimension-five derivative matter-gravity couplings
contribute to $a_i$, removing the logarithmic terms, if we use the prescription
$\int\d^4p 1\propto ({\rm mass})^4$.

There is another issue to take care of. We want to require any renormalized
cosmological constant to vanish. This can be done by properly choosing
$\Lambda_0$ without a serious fine tuning as follows: First of all, we take the
scale $M$ to be small for weak gravity.  We need $\mu$ and $m$ to be large
enough as we want $V_{\rm eff}$ is larger than all the rest suppressed
nonrenormalizable terms. Then we can show that there is $\Lambda_0$ to satisfy
this requirement as long as $\mu$ and $m$ are of the same order. This
determines the numerical relation between $m$ and $\mu$ in terms of $\lambda$,
$M$ and $\Lambda_0$.  Actual computation requires to locate the true vacuum
first, but unfortunately the true vacuum cannot be located in this linearized
gravity. It only indicates the true vacuum, which should exist because one can
always make the potential positive asymptotically, is probably
located in the strong gravity regime. 

Thus, at least in semi-classical analysis, it certainly indicates that the
original vacuum is no longer stable simply because $\hat{h}$ and
$\hat{\gamma}_{\mu\nu}$ behave differently now. More precisely, if  $a_1 \neq
\half\Lambda$, it indicates the vacuum instability in the linearized gravity.
In other words, from the curved spacetime point of view, $\eta_{\mu\nu} +\kappa
h_{\mu\nu}$ does not behave like a metric. Thus, from the generation of this
effective potential, we can anticipate that the symmetry,
eqs.(\ref{eqtf1})(\ref{eqtf2}), is probably spontaneously broken. The remaining
symmetry is nothing but the SDiff symmetry, which can be seen easily from
eq.(\ref{esdiff}).  Once we accept that the symmetry breaking occurs, we can
use eq.(\ref{e7}) to compute the mass of the dilaton consistently because the
effective potential is still SDiff-invariant. 

%\newsection{Diagonalization of the Kinetic Energy}

%\noindent
$\CL_{\hat{h}}$ contains a mixed term of $\hat{h}$ and $\hat{\gamma}_{\mu\nu}$. 
To obtain a system in which $\hat{h}$ and $\hat{\gamma}_{\mu\nu}$ are
completely independent, we need to get rid of such a term. In the usual
linearized gravity this term is removed by a gauge fixing in the Newtonian
limit. Here we can remove this mixing term by incorporating the string dilaton.
In fact we need to introduce the string dilaton if  the linearized gravity  is
to be inherited from a well-defined quantum gravity like string theory. The
string dilaton is defined in the string frame such that it couples to the
world-sheet curvature scalar\cite{ftdil}. Therefore the dilaton in the Einstein
frame appears as a combination of the conformal dilaton and the string dilaton.
Such manipulation for the linearized gravity in the string context can be found,
for example, in \cite{polch}\cite{lanelson}\cite{zwie}. This is due to the
fact that the natural string frame and the Einstein frame are different. 

This leads us to use the following Lagrangian for the string dilaton $\phi$:  
\beq
\label{edil}
\CL_\phi = a \pa_\mu\phi\pa^\mu\phi 
+{\textstyle{3\over 4}}\pa_\mu\phi\pa^\mu \hat{h} - 
\pa^\mu\phi\pa^\nu\hat{\gamma}_{\mu\nu},
\eeq
where $a$ is a constant. $a=2$ corresponds to the string effective action for 
conformal backgrounds written in the string frame. Thus $\phi$ in this case 
actually denotes the string dilaton. We shall however
leave $a$ arbitrary for future purposes.

Now let us introduce a field redefinition
\beq
\label{eq3}
{h}_{\mu\nu} = 2\eta_{\mu\nu}\phi + \tilde{h}_{\mu\nu}
+\kappa(2\phi\tilde{h}_{\mu\nu} + 2 \eta_{\mu\nu}\phi^2).
\eeq
This field redefinition is nothing but the linearized version of  $g_{\mu\nu} =
\e^{2\kappa \phi}\tilde{g}_{\mu\nu}$ and  makes sense, despite that $\phi$
transforms like a scalar. It corresponds to mixing of the string dilaton $\phi$
and the conformal dilaton $\hat{h}$.  One can also easily check out that
$\tilde{h}_{\mu\nu}$ is in fact in  the Einstein frame. This field redefinition
only affects the conformal dilaton so that  $\hat{\gamma}_{\mu\nu} =
\hat{\tilde{\gamma}}_{\mu\nu}$.  We then obtain the identity
\beq
\label{eq4}
\CL_0 +\CL_{\hat{h}} + \CL_\phi = \CL_0 + \CL_{\hat{\tilde{h}}} 
+ (3+a)\pa_\mu\phi\pa^\mu\phi.
\eeq
If we choose $\hat{\tilde{h}}\equiv \tilde{h}
-\half\kappa\tilde{h}^{\alpha\beta}\tilde{h}_{\alpha\beta} = 0$, 
$\CL_{\hat{\tilde{h}}}$ drops out so that we can successfully diagonalize  the
kinetic energy terms in the Lagrangian. In fact this corresponds to choosing
the  traceless gauge in the Einstein frame and $\hat{h} = 8\phi$.  
$\hat{\tilde{h}} =0$ can be chosen because of the presence of $\phi$ degrees of
freedom. In this context, $\hat{\tilde{h}}$ takes the role of the
would-be-Goldstone boson.

Thus we obtain the SDiff-invariant effective Lagrangian 
\beq
\label{eq5}
\CL_{\rm eff}(\tilde\gamma, \phi) = \CL_0(\tilde{\gamma}) 
+ {(a+3)}\pa_\mu \phi\pa^\mu \phi -V_{\rm eff}(8\phi)
\eeq
The vev of $\phi$ in principle can be determined by minimizing 
$V_{\rm eff}(8\phi)$.

%\newsection{Comparison to the Newtonian Gravity}

%\noindent
We often hesitate to abandon the manifest Diff-invariance because we  are
afraid that it may lead to inconsistency of a theory. However, the lesson we
learned from gauge theory is that, as long as a local symmetry is not
explicitly broken, we can have a consistent theory without a manifest local
symmetry. Here we are in a similar situation. The Diff symmetry is treated the
same way as any local internal symmetry and is (probably) spontaneously broken
because the manifestly symmetric vacuum has been destabilized. The conformal
dilaton and the graviton do not behave in the same way. As a first step to
check the consistency, we need to ask if the weak gravity limit could be in
fact governed by such a theory in a low energy scale. The only criterion we
need to satisfy is the existence of the correct Newtonian limit in this
framework.

From eq.(\ref{eq5}) we can derive the equations of motion in the string frame
for $\gamma_{\mu\nu} = \hat{\gamma}_{\mu\nu}(\kappa = 0)$ and $h$ as
\beq
\label{eq6}
{\textstyle{1\over 4}}\pa^\mu\pa_\mu{\gamma}_{\alpha\beta} + 
{(a+3)\over 32}\eta_{\alpha\beta}\pa^\mu\pa_\mu h +\eta_{\alpha\beta} 
V'_{\rm eff}(h)= -\half T_{\alpha\beta}
\eeq
with further gauge fixing $\pa^\mu\gamma_{\mu\nu} = 0$.
Eq.(\ref{eq6}) factorizes into the trace part and the rest. The trace part 
depends only on the conformal dilaton and reads
\beq
\label{eq7}
{(a+3)\over 8}\pa^\mu\pa_\mu h + 4 V'_{\rm eff}(h) = 
-\half T \equiv -\half T^\mu_{\ \mu}.
\eeq
The rest takes a familiar form
\beq
\label{eq8}
\pa^\mu\pa_\mu\gamma_{\alpha\beta} = -2 \tilde{T}_{\alpha\beta},
\eeq
where $\tilde{T}_{\alpha\beta} \equiv {T}_{\alpha\beta} 
-{1\over 4}\eta_{\alpha\beta} T$. This clearly shows that the conformal dilaton
and the graviton behave independently.

To take the Newtonian limit we first transform into the Einstein frame.
Eq.(\ref{eq8}) remains the same, but the trace part changes. It is more 
instructive if we rewrite the trace part in terms of $\phi$ so that
\beq
\label{eq9}
(a+3)\pa^\mu\pa_\mu\phi + \half\tilde{V}'_{\rm eff}(\phi) = -\half T,
\eeq
where $\tilde{V}'_{\rm eff}(\phi) \equiv 8 V'_{\rm eff}(h)$.
Then, taking $T_{00} = \rho$, $T_{io} = 0 = T_{ij}$, one can easily see that
eq.(\ref{eq8}) leads to the correct Newtonian limit in the Einstein frame. 
We also need to turn off any quantum effect so that we set $V_{\rm eff} = 0$.
Then $a = -5$ is required to be consistent at the Newtonian
limit so that $\nabla^2\phi = -\rho/4$. Thus spontaneous Diff symmetry 
breaking can be consistent with the Newtonian limit.

Having $a=-5$ is rather unpleasant because the string effective action with a
conformal background does not satisfy this condition. But, this does not
necessarily mean that such symmetry breaking does not occur in string theory.
Furthermore, we do not really expect the perturbative string action with a
conformal background will describe the physics at the low energy limit because
of the strong coupling  nature of the string theory\cite{ds}\cite{banksdine}.
If we wish, 
we could always rescale the stress-energy tensor to meet the requirement.

It is also important to point out that $\phi$ is not the Brans-Dicke 
field\cite{weinbergb}, which provides additional gravitational degrees of
freedom. $\phi$ simply replaces the trace part of the graviton with additional
dynamics at higher energy scale.

%\newsection{Conclusion}

%\noindent
Although we are not able to show the symmetry breaking explicitly because of
our inability to precisely locate the true vacuum, we have at least shown the
instability of the original Diff-symmetric vacuum. Therefore, the dilaton gets
a vev and presumably there is a true vacuum because the potential is
asymptotically positive.

Let us recapture the essence of the spontaneous symmetry breaking.  At first
sight, it looks quite different from gauge symmetry breaking, yet it has
certain resemblance. Although no separate symmetry breaking sector is
introduced as in the case of dynamical symmetry breaking, but the conformal 
dilaton takes its role and develops a vev. $\hat{\tilde{h}}$ takes the role 
of the would-be-Goldstone boson eaten by  $\phi$ and, as a result, $\phi$ 
(presumably) becomes massive. One may think that a gauge field disappears to 
provide a mass to a scalar field, but it is not.  In fact, $\phi$ in the 
Einstein frame is equivalent to $\hat{h}$ in the string frame so that one can 
think of $\phi$ disappearing on behalf of $\hat{h}$ in $\hat{\tilde{h}} = 0$ 
gauge.  We have simply renamed fields in terms of a field redefinition.  
Therefore, we anticipate radiative spontaneous Diff symmetry breaking and the 
remaining symmetry is the SDiff symmetry. 

The true vacuum is likely to be located in the strong gravity region, 
indicating the breaking of the Diff symmetry down to the SDiff symmetry occurs 
at much higher energy scale.  We need a strong gravity formulation to locate
the true vacuum precisely. We hope further investigation along the line of
ref.\cite{sdiffgr} in 4-d theory would shed some light on the location of the
true vacuum and  a rigorous proof of such a symmetry breaking. A rigorous proof
needs to show two things: quantization of the linear graviton and gauge
invariant computation of higher order terms. We plan to address these issues
elsewhere.

This also further supports the conjecture that generation of the nontrivial
dilaton potential in string theory might necessarily require spontaneous
breaking of Diff-invariance down to SDiff-invariance\cite{sdiffgr}.

%\nopagebreak

{\renewcommand{\Large}{\large}

}

\end{document}